# Measuring the Learning from Two-Stage Collaborative Group Exams

Joss Ives

*Dept. of Physics & Astronomy, University of British Columbia, 6224 Agricultural Road, Vancouver, BC V6T 1Z1*

**Abstract:** A two-stage collaborative exam is one in which students first complete the exam individually, and then complete the same or similar exam in collaborative groups immediately afterward. To quantify the learning effect from the group component of these two-stage exams in an introductory Physics course, a randomized crossover design was used where each student participated in both the treatment and control groups. For each of the two two-stage collaborative group midterm exams, questions were designed to form matched near-transfer pairs with questions on an end-of-term diagnostic which was used as a learning test. For diagnostic test questions paired with questions from the first midterm, which took place six to seven weeks before the diagnostic test, an analysis using a mixed-effects logistic regression found no significant differences in diagnostic-test performance between the control and treatment group. For diagnostic test questions paired with questions from the second midterm, which took place one to two weeks prior to the diagnostic test, the treatment group performed significantly higher on the diagnostic-test than control.

## INTRODUCTION

A two-stage collaborative group exam is one in which students first write the exam individually, and then immediately after those individual exams are collected, the students write the collaborative group portion of the exam, typically in groups of three or four. Although implementations vary, in this study we used a format where each group was provided only a single copy of the group exam and the majority of the questions on the collaborative group exam were the same as those that appeared on the individual portion of the exam. Exam grades were calculated using a weighting of 85% from the individual exam grade and 15% from the group exam grade, with the exception that the weighting would be 100% from the individual exam grade if it was higher than the group exam grade.

The collaborative group exam portion of these two-stage exams provides students with feedback at a time when they are intensely engaged with the material [1] and it may offer similar learning benefits to those provided by Peer Instruction [2], but with an even higher level of student engagement. In addition to these direct learning benefits, it has been shown that two-stage exams have many affective benefits [3,4] (see also Gilley & Clarkston [5] for a concise summary of affective benefits in the collaborative testing literature).

Previous studies [5,6,7], which have shown that there was improved retention after collaborative testing, used the same questions for the retest as were used in the initial test. Other studies (see Leight *et al.* [4] and the cited studies therein) that found no improved retention ranged from using the same individual questions being retested to sets of retest questions that matched the initial questions only in broad topic.

Most of the studies discussed thus far have not controlled for time on task, thus it could be argued that the observed positive retention results were due to some unknown combination of the group exams themselves and the enhanced retention that comes from a well-established phenomenon known as the testing effect [8]. However, similar results from a study that controlled for time on task [5] and one that did not [7] suggest that the major contribution to the improved

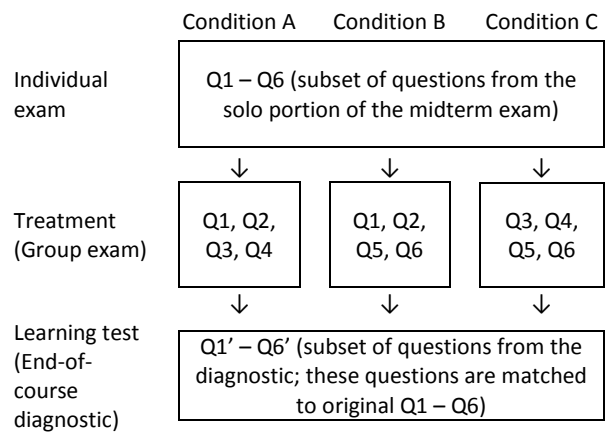

**FIGURE 1.** Flow diagram of the experimental design for each of the two-stage collaborative midterms. The group test followed immediately after the individual test.

retention is from the group exam intervention and not the additional time on task.

In the study presented in this manuscript, pairs of near-transfer [9] questions were used for the initial test and retest to remove the confounding factor of question recall from the measure of learning.

## METHODS

### Experimental Design

Participants were students from an introductory calculus-based fluids, waves and energy course, offered during the January, 2014 term. The following describes the experimental design for the first midterm exam ($n = 679$), which was also repeated for the second midterm exam ($n = 673$). A randomized crossover design was used, where each participant was in both the control and treatment groups (see Fig. 1). Six questions from each exam were developed to match six questions on the end-of-course conceptual diagnostic, in terms of near-transfer application of a given concept. This diagnostic test has been developed specifically for this course and has been in development since 2012. The midterm questions that were designed as the matched-pair partners were typical of the multiple-choice exam questions administered in the course, but were also designed specifically to be near-transfer matched partners with the previously developed questions from the diagnostic. Three versions of the group exam were created, each missing two of the six questions found on the individual exam. These three group exams were distributed randomly to the self-organized groups when they started the group-exam portion of the midterm, thus each student was randomly assigned to condition A, B, or C (see Fig. 1). Although there were three sections of the course running at the same time, each with different instructors, the midterms were written at common times and the randomization of conditions was done the same way within each section. For each condition, a participant answered four of the six questions on both the individual and group portions of the exam (treatment) and two of the six questions only on the individual portion of the exam (control). This crossover design was felt to be one which was very fair to the students from the perspective of each student in the course having been provided equal opportunity to participate in and learn from the collaborative-group portion of the exam.

The end-of-course conceptual diagnostic was administered during the each student's last laboratory session of the term, which depending on a student's scheduled lab section, took place six to seven weeks after the first midterm and one to two weeks after the second midterm. Students were offered a participation bonus of 1% toward their final grade for completing the diagnostic at the start and the end of the term.

### Question Validation

Three aspects of question validation will be discussed: validation of the question wording, question reliability through classical test theory item analysis, and rating the matched questions pairs for how well they target the same application of a given concept.

The questions from the end-of-course diagnostic are all in the mid-to-late stages of an iterative validation process [10], which has consisted of feedback from local experts (primarily course instructors), talk-aloud interviews with students whom have previously taken the course, and various classical test theory item analyses.

The midterm exam questions were developed collaboratively between the four course instructors (the manuscript's author being one of these course instructors) to be near-transfer matched-pair partners of the twelve diagnostic test questions used in this study. As is the common practice in this course, the midterm questions were reviewed for clarity by the majority of the graduate teaching assistants assigned to the course before being administered to the students.

To quantify how well the questions in a matched pair (one from the midterm and one from the learning test) target the same application of a given concept, seven content experts were asked to rate each question pair on a five-point scale where a rating of five meant the questions target the same application of the same concept, three meant the questions targeted different applications of the same concepts, and one meant the questions targeted completely different concepts. These ratings are shown in Table I. All the question pairs were rated as a three or higher and two-thirds were rated as a four or higher. There is insufficient statistical power to determine if question pairs from midterm two (Q7 – Q12) differ significantly in similarity rating to question pairs from midterm one (Q1 – Q6). The four pairs of questions from midterm 2 with similarity ratings of 4.86 and above represent questions that were nearly identical between the midterm and learning test. This point will be discussed further in the conclusions.

To determine how well each question discriminates between high- and low-performing students (highest 21% and lowest 21% of scores on the 11 other learning test questions, respectively), the item discrimination index [11], $D$, was used. The highest possible value, $D = 1$, would indicate that all the high-performing students answered the question correctly and none of the low-performing students answered the question correctly.

**TABLE I.** Measures used for question validation and class performance for Q1 – Q6 (Q7 – Q12) represent the six questions from the first (second) midterm and the associated diagnostic test questions.

| Midterm and Question Num. | Similarity Rating (SD) | Exam Questions | | Learning Test Questions | | |
|---|---|---|---|---|---|---|
| | | Fraction Correct | Discrimination Index, $D$ | Fraction Correct, Control | Fraction Correct, Treatment | Discrimination Index, $D$ |
| MT1, Q1 | 3.29 (1.11) | .453 | .341 | .507 (N=205) | .475 (N=474) | .270 |
| MT1, Q2 | 4.00 (0.58) | .474 | .324 | .449 (N=274) | .486 (N=405) | .315 |
| MT1, Q3 | 4.71 (0.76) | .636 | .440 | .434 (N=205) | .384 (N=474) | .465 |
| MT1, Q4 | 4.57 (0.53) | .744 | .403 | .514 (N=274) | .551 (N=405) | .387 |
| MT1, Q5 | 3.14 (1.07) | .610 | .490 | .440 (N=200) | .397 (N=479) | .345 |
| MT1, Q6 | 4.28 (0.49) | .820 | .335 | .500 (N=200) | .461 (N=479) | .405 |
| MT2, Q7 | 3.71 (1.11) | .841 | .231 | .683 (N=218) | .741 (N=455) | .275 |
| MT2, Q8 | 3.86 (1.46) | .634 | .370 | .637 (N=218) | .646 (N=455) | .200 |
| MT2, Q9 | 4.86 (0.38) | .837 | .167 | .703 (N=236) | .721 (N=437) | .385 |
| MT2, Q10 | 4.86 (0.38) | .626 | .305 | .737 (N=236) | .705 (N=437) | .432 |
| MT2, Q11 | 5.00 (0.00) | .691 | .399 | .571 (N=219) | .654 (N=454) | .464 |
| MT2, Q12 | 4.86 (0.38) | .284 | .402 | .342 (N=219) | .425 (N=454) | .500 |

The lowest possible value, $D = -1$, would indicate the opposite, with all low-performing students answering correctly and none of the high-performing students answering correctly. Typically, an item with $D \geq 0.3$ is considered to have good discrimination. Although three of the learning test questions fall below this threshold, an analysis detailed later in the manuscript found that removing these questions from the analysis had no significant impact on the findings of this study.

## ANALYSIS AND RESULTS

The analysis, which will be described below, found that there were no differences between control and treatment for the diagnostic test questions related to midterm one, but that the control group outperformed the treatment group for the learning-test questions related to midterm two.

To examine the impact of the treatment condition, the following mixed-effects logistic regression was used,

$$\text{Log\_odds}(\text{Retest\_success}_{ijk}) = \beta_0 + \beta_{1j} \times \text{Group}_j + \beta_{2k} \times \text{Question}_k + \beta_3 \times \text{Treatment} + \varepsilon_i, \quad (1)$$

where Retest_success$_{ijk}$ is the (binary) success of Student$_i$ from Group$_j$ on Question$_k$ on the learning test, Group$_j$ is a categorical variable representing to which condition group (A, B or C) the student was randomly assigned, Question$_k$ is a categorical variable representing question number and accounted for differences in question difficulty, and $\varepsilon_i$ is a random intercept for Student$_i$ which accounts for differences in student ability. A positive $\beta_3$ would indicate that the group exams had a positive effect on learning-test success.

For diagnostic test questions associated with midterm one, the fit of the model to the data was good, $\chi^2(9)=111.8$, $p<.001$, and correctly predicted 72% of the cases. It was found that treatment had no statistically significant predictive power for the diagnostic test questions associated with midterm one, $p(\beta_3) = .40$.

For diagnostic test questions associated with midterm two, the fit of the model to the data was good, $\chi^2(9)=225.2$, $p<.001$, and correctly predicted 77% of the cases. It was found that treatment predicted diagnostic test success ($\beta_3 = .198$, $SE = .079$, $p = .012$). Expressed as an odds ratio, the odds of answering a question correctly on the diagnostic test (versus not answering it correctly) increased by a factor 1.22 (95% CI [1.04, 1.42]) for those in the treatment condition as compared to the control condition.

To ensure that the results were not being skewed by the question pairs that were doing the most poor job of testing the same application of the same concept, the logistic regression analyses were repeated after removing the four question pairs having similarity ratings below 4.0 (see Table I), three of which are the questions with $D < 0.3$. A similarity rating of 4.0 is the threshold below which the question pair is considered to be closer to "different applications of the same concept" than to "same application of the same concept." Removing these questions did not improve the quality of the fit of the model to the data nor did it change the significance level of $\beta_3$ for the set of learning test questions for either midterm.

The mixed-effects logistic regression analysis found that the treatment had a positive effect on diagnostic test performance when the time between midterm and learning test was one to two weeks, but had no statistically significant effect when the time between midterm and learning test was six to seven weeks.

## SUMMARY AND CONCLUSIONS

For the diagnostic test questions associated with midterm one, participation in the collaborative group exam six to seven weeks earlier was not a statistically significant predictor of success. In contrast, participation in the collaborative group exam one to two weeks earlier than the diagnostic test was a statistically significant predictor of success on the questions associated with midterm two.

The results of this study offer mixed evidence related to improved student learning through the use of two-stage collaborative group exams. Although improved student learning was shown for the treatment group for the diagnostic test questions associated with midterm two, the results were nowhere near as dramatic as those found in other studies which used an effectively similar design [5,7] and also had similar times (relative to the diagnostic test questions associated with midterm two) between the initial test and the learning test.

Given previous results, the lack of evidence for improved student learning for the diagnostic test questions associated with midterm one is surprising. Given the size of the positive effect observed for the questions associated with midterm two, a likely explanation for the null result for the diagnostic test questions associated with midterm one is that it is due to exponential decay in memory [12]. A future study that would isolate this time-delay effect would be to subdivide the population into groups that take the learning test at different times relative to the initial collaborative exam, with times on the scale of a small number of days through the times used in this study. Additional improvements to this study design include use of sets of questions that have undergone even more thorough validation and question-pair matching.

It is also plausible that the difference in effects between the questions associated with midterm one and those associated with midterm two are due to midterm two having four of the six question pairs with extremely high similarity ratings (4.86 out of 5 or higher), where none of the question pairs associated with midterm one were this high. A future study, to perform alongside or after one that isolated the time-delay effect, is one that isolated the effect of question similarity.

The study presented in this manuscript was designed only to measure impacts on learning at the scale of the specific questions used in this study and it fails to explore the beneficial affective impacts. Because the collaborative group exam structure is very similar to Peer Instruction and other group-work activities used throughout the course, the group exams generate student buy-in for the pedagogical techniques used in the course and vice-versa. The group exams also model many positive metacognitive skills and study behaviors, such as reviewing graded exams when returned, defending an answer for the purpose of clarifying understanding or looking at a problem from multiple points of view.

## ACKNOWLEDGMENTS

We would like to thank Hao Luo from UBC's Statistical Consulting and Research Laboratory (SCARL) for consultation on the analysis. We would also like to thank many members of the Physics 101 instructional team and the Physics and Astronomy Education Research (PHASER) group for their contributions to question design, refinement and validation. Additional thanks to Natasha Holmes and Georg Reiger for valuable feedback on the manuscript, and to Brett Gilley and Ido Roll for guidance and advice on experimental design and analysis.